\documentclass[12pt]{article}

\usepackage{graphics,amssymb,epsfig,float}
\usepackage[usenames,dvips]{color}
\usepackage{graphicx}
\usepackage{epsfig}
\usepackage{rotating}
\usepackage{dcolumn}
\usepackage{bm}
\usepackage{cite}
\usepackage{amsmath}

\def\lsim{\;\raise0.3ex\hbox{$<$\kern-0.75em\raise-1.1ex\hbox{$\sim$}}\;}
\def\gsim{\;\raise0.3ex\hbox{$>$\kern-0.75em\raise-1.1ex\hbox{$\sim$}}\;}
\def\beq{\begin{equation}}   \def\eeq{\end{equation}}
\def\ba{\begin{array}}       \def\ea{\end{array}}
\def\bea{\begin{eqnarray}}   \def\eea{\end{eqnarray}}
\def\nn{\nonumber}

\def\k{\kappa}
\def\l{\lambda}

\begin{document}

\begin{titlepage}
\begin{flushright}
LPT Orsay 11-38 \\ 
TTP 11-13
\end{flushright}

\begin{center}
\vspace{1cm}
{\Large\bf Reduced branching ratio for $H\to AA \to 4\,\tau$\\ from $A -
\eta_b$ mixing} \\
\vspace{1cm}

{\bf Florian Domingo$^1$ and Ulrich Ellwanger$^2$}\\
\vspace*{.5cm}
$^1$ Institut f\"ur Theoretische Teilchenphysik, KIT, Universit\"at
Karlsruhe,\\ D-76128 Karlsruhe, Germany\\
$^2$  Laboratoire de Physique Th\'eorique, UMR 8627, CNRS and
Universit\'e de Paris--Sud,\\
B\^at. 210, F-91405 Orsay, France \\

\end{center}

\vspace{1cm}

\begin{abstract}
Models with an extended Higgs sector, as the NMSSM, allow for scenarios
where the Standard Model-like CP-even Higgs boson $H$ decays dominantly
as $H \to AA \to 4\,\tau$ where $A$ is a light CP-odd Higgs boson. Tight
constraints on this scenario in the form of lower bounds on $M_H$ have
recently been published by the ALEPH group. We show that, due to $A -
\eta_b$ mixing, the branching ratio $H\to AA \to 4\,\tau$ is strongly
reduced for $M_A$ in the range $9 - 10.5$~GeV. This is the range of
$M_A$ in which the tension between the observed $\eta_b(1S)$ mass and
its prediction based on QCD can be resolved due to mixing, and which is
thus still consistent with a light CP-even Higgs boson $H$ satisfying
LEP constraints with a mass well below 114~GeV. This result is
practically independent from the coupling of $A$ to $b$ quarks.
\end{abstract}

\end{titlepage}

\section{Introduction}

One of the main goals of the Large Hadron Collider (LHC) is the
detection of the Higgs boson, or of at least one of several Higgs bosons
if corresponding extensions of the Standard Model (SM) are realized in
nature. These searches depend crucially on the Higgs production cross
sections and the Higgs decays.

In the case of the SM, the production cross sections and decay branching
ratios are quite well known as functions of the still unknown Higgs mass
\cite{Djouadi:2005gi}. In the Minimal Supersymmetric Standard Model
(MSSM) with its extended Higgs sector, these quantities have been
studied as well and it seems that at least one of the Higgs bosons
cannot be missed at the LHC \cite{Djouadi:2005gj}. There exist, however,
well motivated scenarios with somewhat more extended Higgs sectors, as
the Next-to-Minimal Supersymmetric Standard Model (NMSSM, see
\cite{Maniatis:2009re,Ellwanger:2009dp} for recent reviews), where the
Higgs decays can differ strongly from both the SM and the MSSM. It is
very important to be aware of the possibility of such unconventional
Higgs decays; the absence of a signal in standard Higgs search
channels may otherwise be completely misinterpreted.

The Higgs sector of the NMSSM consists of two SU(2) doublets $H_u$ and
$H_d$ (as in the MSSM), and one additional gauge singlet $S$. Due to its
coupling $\l S H_u H_d$ in the superpotential, a vacuum expectation
value (vev) $s$ of $S$ generates a supersymmetric mass term
$\mu_\mathrm{eff}=\l s$ for $H_u$ and $H_d$. Since $s$ and hence
$\mu_\mathrm{eff}$ are naturally of the order of the soft Susy breaking
terms $\sim M_\mathrm{Susy}$, this solves the so-called $\mu$-problem of
the MSSM \cite{Kim:1983dt}. (This remains true in the limit $\l,\ \k\to
0$, where $\k$ is the singlet self-coupling in the superpotential,
leading to $s\sim M_\mathrm{Susy}/\k$, but $\mu_\mathrm{eff} \sim
(\l/\k)M_\mathrm{Susy} \sim M_\mathrm{Susy}$.) Furthermore, in its
simplest $Z_3$ invariant version, the superpotential of the NMSSM is
scale invariant; it is in fact the simplest phenomenologically
acceptable supersymmetric extension of the SM with this property.

The physical neutral Higgs sector in the NMSSM consists of 3 CP even and
2 CP odd states. (Here we do not consider the possibility of CP
violation in the Higgs sector.) In general, these states are mixtures of
the corresponding CP even or CP odd components of $H_u$, $H_d$ and $S$,
without the CP odd Goldstone boson swallowed by the massive $Z$ boson.
Often, one of the CP even states is SM like, i.e. with similar couplings
to gauge bosons as the SM Higgs boson (but with possibly enhanced
couplings to quarks and leptons), with a mass bounded from above by
$\sim 140$~GeV \cite{Ellwanger:2006rm}. At first sight, the detection at
the LHC of this Higgs boson -- denoted subsequently by $H$ for
simplicity -- seems to be guaranteed, given the lower LEP bound of $\sim
114$~GeV on masses of Higgs bosons with SM like couplings to the $Z$
boson and SM like decays.

However, the lighter of the two CP odd states (denoted by $A_1$) could
have a mass $M_{A_1}$ below half of the mass $M_H$ of $H$
\cite{Dobrescu:2000jt, Dobrescu:2000yn}. Then, $H$ would decay
dominantly as $H \to A_1 A_1$, since this coupling is typically larger
than the coupling of $H$ to $b$ quarks \cite{Dobrescu:2000jt,
Dobrescu:2000yn}. Such a decay of $H$ would have important consequences
both for lower bounds on its mass from searches at LEP, and for its
detection at the LHC. Now the $H$ final decay products depend on
$M_{A_1}$: for $M_{A_1}\gsim 10.5$~GeV, they consist mainly of 4
$b$~quarks (with some $2\,b+2\,\tau$ admixture), whereas for $3.5\
\mathrm{GeV} \lsim M_{A_1}\lsim 10.5$~GeV, they consist mainly of 4
$\tau$~leptons (with some small $2\,\tau+2\,\mu$ admixture). In fact, $H
\to 4\,b$ decays have also been searched for by OPAL and DELPHI at LEP
\cite{Abbiendi:2004ww,  Abdallah:2004wy} implying $M_H \gsim 110$~GeV if
$H$ has SM like couplings to the $Z$ boson \cite{Schael:2006cr}. On the
other hand, LEP constraints on $H \to 4\,\tau$ decays were relatively
weak, allowing for $M_H$ as low as $\sim 90$~GeV \cite{Schael:2006cr}.

This led to the scenario advocated in \cite{Dermisek:2005ar,
Dermisek:2005gg,Dermisek:2006wr, Dermisek:2007yt} (see also
\cite{Gunion:2011hs}) with $M_H \lsim 110$~GeV, $M_{A_1}\lsim 10.5$~GeV,
a dominant (but not exclusive) decay $H \to A_1 A_1 \to 4$~leptons and a
low finetuning among the soft Susy breaking parameters due to the
relatively low mass of $H$. A remaining small branching ratio for $H\to
2\,b$ could explain the $2\,\sigma$ excess observed in this channel for
$M_H \sim 100$~GeV~\cite{Schael:2006cr, Dermisek:2005gg}.

The final state $H \to 4\,\tau$ has recently been reanalysed by the
ALEPH group \cite{Schael:2010aw} implying upper bounds on $\xi^2 =
\frac{\sigma(e^+ e^- \to ZH)}{\sigma_\mathrm{SM}(e^+ e^- \to ZH)} \times
BR(H \to 2\, A_1) \times BR(A_1\to \tau^+\, \tau^-)^2$ as function
of $M_H$ and $M_{A_1}$. These bounds seem to impose strong
constraints on the above scenario, unless $\sigma(e^+ e^- \to ZH)$
and/or the $BR(H \to 2\, A_1)$ and/or the $BR(A_1\to \tau^+\, \tau^-)$
are smaller than naively expected \cite{Dermisek:2010mg}.

A light CP odd scalar $A_1$ would also have important consequences for
the physics of $b\bar{b}$ bound states. These effects depend on the
coupling of $A_1$ to $b$~quarks. Normalized relative to the coupling of
the SM Higgs boson, the coupling of $A_1$ to $b$~quarks is given by
$X_d$ with
\beq\label{eq:1}
X_d = \cos\theta_{A} \tan\beta\, ,
\eeq
where $\cos\theta_{A}$ denotes the SU(2) doublet component of $A_1$,
and $\tan\beta$ is the usual ratio of Higgs vevs $v_u/v_d$. For
$\tan\beta$ much larger than 1, $X_d$ could satisfy $X_d \gg 1$ as well.
($X_d$ is simultaneously the coupling of $A_1$ to leptons normalized
relative to the coupling of the SM Higgs boson.)

In fact the relation (\ref{eq:1}) is valid for $A_1$ in any extension of
the SM with two Higgs doublets $H_u$ (coupling exclusively to up-type
quarks) and $H_d$ (coupling exclusively to down-type quarks and
leptons), but arbitrary singlets. Our subsequent results depend only on
$M_{A_1}$ and $X_d$, and are valid for any such models. In the NMSSM, a
light CP odd scalar $A_1$ can play the role of a pseudo Goldstone boson
of an approximate R- or Peccei-Quinn symmetry \cite{Dobrescu:2000jt,
Dobrescu:2000yn}. Then, however, one always has $\cos\theta_{A} \sim
1/\tan\beta$ \cite{Ellwanger:2009dp} and hence $X_d \lsim 1$.

Since the pseudoscalar $b\bar{b}$ bound states $\eta_b(nS)$ have the
same quantum numbers as a CP odd Higgs $A_1$, the states $\eta_b(nS)$
and $A_1$ can mix \cite{Drees:1989du,Fullana:2007uq,Domingo:2008rr,
Domingo:2009tb} with important consequences both for the mass spectrum
and the decays of the physical eigenstates. A state $\eta_b(1S)$ has
been observed in radiative $\Upsilon(3S)$ and $\Upsilon(2S)$ decays by
BABAR \cite{:2008vj,:2009pz}, with the result that its mass of
$9390.9\pm 3.1$~MeV is below the one expected from most QCD predictions
for the $\Upsilon(1S)-\eta_b(1S)$ hyperfine splitting 
\cite{Recksiegel:2003fm, Kniehl:2003ap,Penin:2009wf}. Indeed, such a
mass shift could be explained by the mixing of $\eta_b(1S)$ with $A_1$
provided $M_{A_1}$ (before mixing) is in the $9.4-10.5$~GeV range
\cite{Domingo:2009tb}.

On the other hand, $A_1$ can be searched for in radiative decays
$\Upsilon(nS)\to \gamma A_1,\ A_1 \to 2$~leptons. (See
\cite{Rashed:2010jp} for a discussion of $\eta_b \to \tau^+ \tau^-$
mediated by $A_1$.) Unsuccessful searches
by CLEO \cite{:2008hs} and BABAR \cite{Aubert:2009cp,Aubert:2009cka}
lead to upper bounds on $X_d$ as function of $M_{A_1}$, which have been
studied in \cite{Dermisek:2006py,Domingo:2008rr,Dermisek:2010mg} for
$M_{A_1} \lsim 9$~GeV where the $\eta_b(nS)-A_1$ mixing is not very
relevant. Notably for $M_{A_1}$ below the $2\,\tau$ threshold, where
$A_1$ has a large branching fraction into two muons, these bounds are
quite strong and imply $X_d \lsim 0.5$. Upper bounds on $X_d$ for $8\
\mathrm{GeV} \lsim M_{A_1} \lsim 10.1$~GeV, including effects from
$\eta_b(nS)-A_1$ mixing, have recently been investigated in
\cite{Domingo:2010am}, implying $X_d \lsim 2\dots 7$ depending on
$M_{A_1}$. (These bounds are consistent with limits from the violation
of lepton universality in inclusive $\Upsilon(nS)$ decays as proposed in
\cite{SanchisLozano:2002pm,SanchisLozano:2003ha, Fullana:2007uq,
SanchisLozano:2006gx} and studied in \cite{Guido:2009xg,
delAmoSanchez:2010bt}.)

Possible $\eta_b(nS)-A_1$ mixings would also affect the ALEPH bounds on
$H \to 2\,A_1\to 4\,\tau$ \cite{Schael:2010aw} in the interesting mass
range $9\ \mathrm{GeV} \lsim M_{A_1} \lsim 10.5$~GeV, since $A_1$
decaying hadronically through its $\eta_b$ components would imply a
different signature. The corresponding consequences for this process
have not been taken into account before; this study is the purpose of
the present paper. In fact, our result is quite dramatic: the ALEPH
bounds imply practically no constraint on the $BR(H \to 2\,A_1)$ in the
corresponding mass range, since the $BR(A_1\to \tau^+ \tau^-)$ tends to
be very small even for small values of $X_d$. The origin of this
phenomenon can easily be understood qualitatively: the width of the
decay $A_1 \to \tau^+ \tau^-$ of the pure state $A_1$, albeit
proportional to $X_d^2$ (which appears also in the coupling of $A_1$ to
$\tau$ leptons), is always much smaller than the hadronic width of the
$\eta_b(nS)$ to hadrons given the present upper bounds on $X_d$. Hence,
even a small admixture of $\eta_b(nS)$ to any physical eigenstate
implies a large hadronic decay width, suppressing the branching ratio of
the physical state into $\tau^+ \tau^-$ and making it very difficult to
detect. For $X_d \lsim 10$ this effect is approximately independent from
$X_d$, since both the width for $A_1 \to \tau^+ \tau^-$ \emph{and} the
$\eta_b(nS)-A_1$ mixing are proportional to $X_d^2$.

In the next Section we study this phenomenon quantitatively, with the
result stated above. In Section~3 we briefly comment on the impact of
our result on future Higgs searches.

\section{The $BR(H \to 4\,\tau)$ in the presence of $A - \eta_b$ mixing}

In this section we consider the mixing of a CP odd Higgs state $A_1$
(denoted by $A$ for simplicity) with the states $\eta_b(1S)$,
$\eta_b(2S)$ and $\eta_b(3S)$ with masses below the $B\bar{B}$
threshold. The mass squared matrix in the basis $\eta_b(1S) -
\eta_b(2S) - \eta_b(3S) - A$ can be written as \cite{Domingo:2009tb}
\beq\label{eq:2}
{\cal M}^2=  
\left(
\begin{array}{cccc}
     m_{\eta_b(1S)}^2 & 0 & 0 & \delta m_1^2\\
     0 & m_{\eta_b(2S)}^2 & 0 &\delta m_2^2\\
     0 & 0 & m_{\eta_b(3S)}^2 & \delta m_3^2\\
     \delta m_1^2 & \delta m_2^2 & \delta m_3^2 & M_A^2
     \end{array}
\right)\; .
\eeq
The off-diagonal elements $\delta m_n^2$ depend on the $\eta_b(nS)$ wave
functions at the origin, and $X_d$ as given in (\ref{eq:1}) multiplied
by  the coupling of a SM like Higgs boson to $b$ quarks
\cite{Drees:1989du,Fullana:2007uq,Domingo:2008rr,Domingo:2009tb}.
Estimating the wave functions at the origin as in
\cite{Fullana:2007uq,Domingo:2008rr,Domingo:2009tb} one obtains
\bea
 &\delta m_1^2&\ \simeq\ 0.14\ \mathrm{GeV}^2\times X_d\;,\nn \\
 &\delta m_2^2&\ \simeq\ 0.11\ \mathrm{GeV}^2\times X_d\;,\nn \\
 &\delta m_3^2&\ \simeq\ 0.10\ \mathrm{GeV}^2\times X_d\;.
 \label{eq:3}
\eea
The errors on these quantities are about 10\%, but our subsequent
results are not sensitive to the precise numerical values. For the
diagonal elements  $m_{\eta_b(nS)}^{2}$ we take \cite{Recksiegel:2003fm}
$m_{\eta_b(2S)} = 10002$~MeV, $m_{\eta_b(3S)} = 10343$~MeV. 
$m_{\eta_b(1S)}^{2}$ is determined, for given $M_A$ and $X_d$, by the
condition that the state with its mass of $\sim 9391$~MeV observed in
radiative $\Upsilon(3S)$ and $\Upsilon(2S)$ decays by BABAR
\cite{:2008vj,:2009pz} must be identified with one of the eigenstates of
${\cal M}^2$. Again, our subsequent results depend only weakly on these
masses.

It is straightforward to diagonalize the mass matrix (\ref{eq:2}). The
4 eigenstates will be denoted by $\eta_i$, which are decomposed into the
unmixed states as
\beq\label{eq:4}
\eta_i = P_{i,1}\;\eta_b(1S)
+ P_{i,2}\;\eta_b(2S)+ P_{i,3}\;\eta_b(3S) + P_{i,4}\;A\;.
\eeq

Both the eigenvalues of the mass matrix (\ref{eq:2}) and the mixing
coefficients $P_{i,j}$ in (\ref{eq:4}) depend on the unknown mass $M_A$.
Let us recall some obvious properties of the eigenvalues and the mixing
coefficients: whenever $M_A$ is far from any of the $m_{\eta_b(nS)}$,
the mixing will be relatively small (but increasing with $X_d$), and
$A$ will be an approximate mass eigenstate. For fixed $X_d$, the closer
$M_A$ is to $m_{\eta_b(nS)}$, the larger the $A - \eta_b(nS)$ mixing
will be, resulting in shifts of the eigenvalues of ${\cal M}^2$ w.r.t.
its diagonal elements.

We recall that the state with a mass of $\sim 9391$~MeV observed by
BABAR must be identified with one of the eigenstates of ${\cal M}^2$.
Independently from the value of the diagonal element $m_{\eta_b(1S)}$ of
${\cal M}^2$, it follows that $M_A$ cannot be arbitrarily close to
$9391$~MeV unless the mixing (and hence $X_d$) tends to zero. This
consideration leads to an upper bound on $X_d$ depending on $M_A$, with
$X_d \to 0$ for $M_A \to 9391$~MeV, and still $X_d \lsim 20$ for $M_A
\sim 10$~GeV or $M_A \sim 8.5$~GeV \cite{Domingo:2008rr}.

Next we turn to the decays of the eigenstates $\eta_i$, starting with
the decays of the states before mixing. $A$ will decay dominantly into
$A \to \tau^+\,\tau^-$, with a partial width $\Gamma_A^{\tau\tau}$ given
by
\beq\label{eq:5}
\Gamma_A^{\tau\tau} = X_d^2 \frac{G_F m_\tau^2 M_A}{4\sqrt{2}\pi}
\sqrt{1-4\frac{m_\tau^2}{M_A^2}}\ \sim\ X_d^2 \times 1.9\cdot 10^{-2}\
\mathrm{MeV} \times \left(\frac{M_A}{10\;\mathrm{GeV}}\right)
\; .
\eeq
We determine the $BR(A \to \tau^+\,\tau^-)$ from NMHDECAY
\cite{Ellwanger:2004xm,Ellwanger:2005dv} inside NMSSMTools
\cite{nmssmtools} (assuming $\tan\beta \sim 5$), which gives $BR(A \to
\tau^+\,\tau^-) \sim 0.9-0.75$ with increasing $M_A$, the remaining $BR$
originating from $A$ decays into $c\bar{c}$ quarks and gluons. (A
smaller $BR(A \to \tau^+\,\tau^-)$, as advocated for some parameter
choices in \cite{Dermisek:2010mg}, would only amplify our subsequent
conclusions.) Hence we take $\Gamma_A^{tot} \sim (1.1 - 1.33) \times
\Gamma_A^{\tau\tau}$.

The states $\eta_b(nS)$ (before mixing) would decay nearly exclusively
into hadrons (like the states $\eta_c(nS)$). Using the formalism
in \cite{Petrelli:1997ge}, the widths of the states $\eta_b (nS)$ can be
estimated from the widths of the corresponding $\Upsilon$ states and the
$\eta_b (nS)$ masses. Subsequently we take $\Gamma_{\eta_b(1S)} =
11.8$~MeV, $\Gamma_{\eta_b(2S)} = 5.4$~MeV and $\Gamma_{\eta_b(3S)} =
3.9$~MeV.  Note that, unless $X_d \gsim 10$, these widths are much
larger than $\Gamma_A^{\tau\tau}$. (We recall that, for $M_A \lsim
10.1$~GeV, $X_d \lsim 2\dots 7$ due to constraints from $\Upsilon(nS)\to
\gamma A_1,\ A_1 \to 2$~leptons \cite{Domingo:2010am}.)

In terms of these widths and the mixing coefficients, the $BR(\eta_i \to
\tau^+\,\tau^-)$ of the eigenstates $\eta_i$ are given by
\cite{Domingo:2009tb}
\beq\label{eq:6}
BR(\eta_i \to \tau^+\,\tau^-) = \frac{P_{i,4}^2 \Gamma_A^{\tau\tau}}
{\displaystyle\left(\sum_{n=1}^3 P_{i,n}^2 \Gamma_{\eta_b(nS)}\right) +
P_{i,4}^2 \Gamma_A^{tot}}\; .
\eeq

Let us consider the state $\eta_i$ with the largest $A$ component, i.e.
the largest coefficient $P_{i,4}^2$. (Since, essentially, $A$ mixes with
just one of the $\eta_b(nS)$ states depending on $M_A$, there exists
always one state with $P_{i,4}^2 \gsim 0.5$.) Its $BR(\eta_i \to
\tau^+\,\tau^-)$ is smaller than $0.9-0.75$ due to the terms $\sim
\Gamma_{\eta_b(nS)}$ in the denominator of (\ref{eq:6}). In fact, even
if $P_{i,n}^2 \ll 1$, these terms are often numerically dominant due to
$\Gamma_{\eta_b(nS)} \gg \Gamma_A^{tot}$, implying a considerable
reduction of the $BR(\eta_i \to \tau^+\,\tau^-)$. For $X_d \lsim 5$, the
result is nearly independent from $X_d$, since $\Gamma_A^{\tau\tau}$ and
$\Gamma_A^{tot}$ as well as $P_{i,n}^2$ are proportional to $X_d^2$, and
$X_d^2$ cancels out.

\begin{figure}[ht!]
\begin{center}
\includegraphics*[width=12cm,height=10cm]{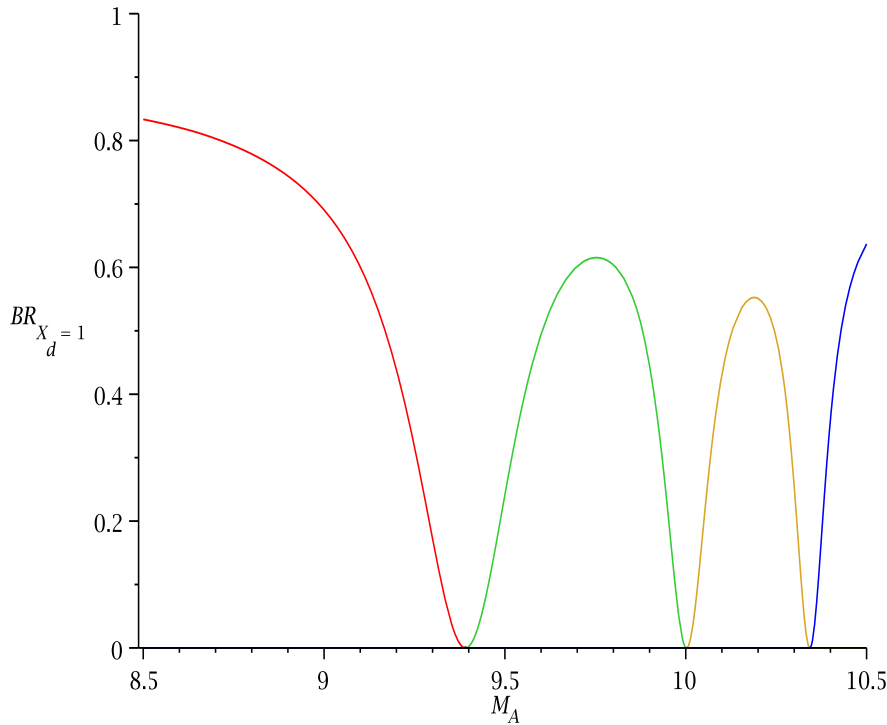}
\caption{The $BR(\eta_i \to \tau^+\,\tau^-)$ for the state $\eta_i$ with
the largest $A$ component as function of $M_A$ for $X_d = 1$. The colors
indicate which state $\eta_i$ is concerned (red$\rightarrow \eta_1$,
green$\rightarrow\eta_2$, brown$\rightarrow\eta_3$, 
blue$\rightarrow\eta_4$).}
\label{fig:1}
\end{center}
\end{figure}

In Fig.~\ref{fig:1} we show the $BR(\eta_i \to \tau^+\,\tau^-)$ for the
state $\eta_i$ with the largest $A$ component as function of $M_A$ for
$X_d = 1$. Depending on $M_A$, this state corresponds to $\eta_1 \dots
\eta_4$, which is indicated by the various colors. For $X_d = 1$, the
mass of this state is practically identical to $M_A$. Usually, the
branching ratios into $\tau^+\,\tau^-$ of the remaining states are
neglibibly small. Note that, whenever $M_A$ is close to any of the
masses $m_{\eta_b(nS)}$, the mixing becomes strong ($P_{i,4}^2 \sim
P_{i,n}^2 \sim 1/2$) leading to $BR(\eta_i \to \tau^+\,\tau^-) \sim
\Gamma_A^{\tau\tau}/\Gamma_{\eta_b(nS)}$, which is very small. (As
stated above, we must have $X_d \to 0$ for $M_A \to 9391$~MeV. This
upper bound is applied to $X_d$ for $M_A \sim 9391$~MeV in
Fig.~\ref{fig:1}, but $X_d = 1$ is used for all other values of $M_A$.)
Remarkably, even if $M_A$ is not close to any of the masses
$m_{\eta_b(nS)}$, the suppression of the $BR(\eta_i \to \tau^+\,\tau^-)$
is still quite strong due to the terms $\sim \Gamma_{\eta_b(nS)}$ in the
denominator of (\ref{eq:6}), and $BR(\eta_i \to \tau^+\,\tau^-) \lsim
0.65$ for any $M_A$ in the range $9 - 10.5$~GeV.

\begin{figure}[ht!]
\begin{center}
\includegraphics*[width=16cm,height=7cm]{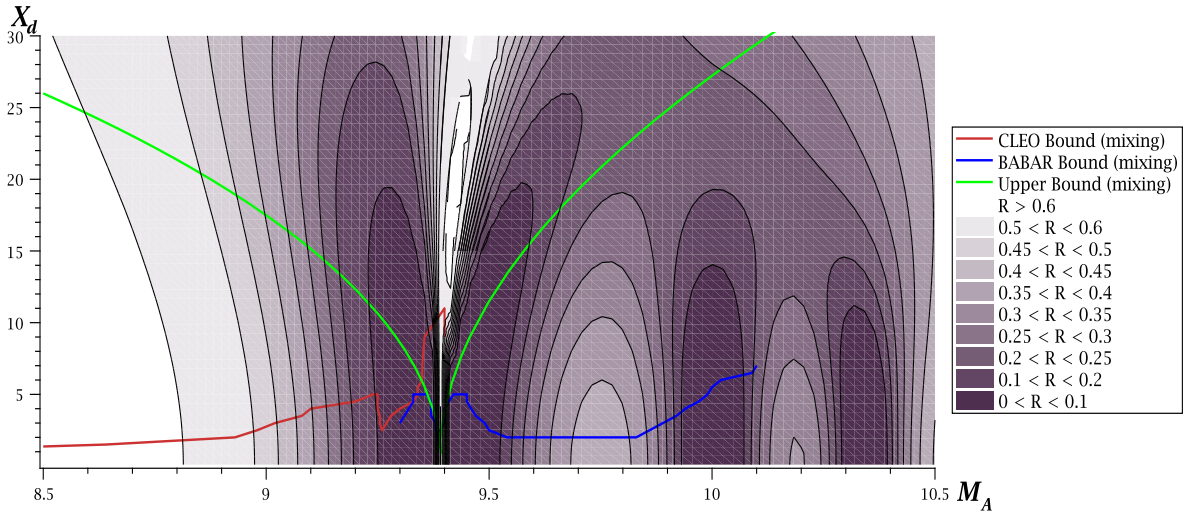}
\caption{The function $R$, defined in (\ref{eq:7}), in the plane $X_d$
vs. $M_A$. Also indicated are upper bounds on $X_d$ from CLEO (red),
BABAR (blue) and from the condition that one eigenstate of the mass
matrix (\ref{eq:2}) has a mass of 9391~MeV with $m_{\eta_{b}(1S)}$
within a reasonable range (green).}
\label{fig:2}
\end{center}
\end{figure}

Finally we have to re-interpret the decay $H\to A A \to 4\,\tau$ in the
presence of $A - \eta_b(nS)$ mixing: now this process corresponds to
$\sum_{i,j=1}^4 (H \to \eta_i\, \eta_j \to 4\,\tau)$. The coupling of
the states $\eta_i$ to $H$ (originating from the coupling of $A$ to $H$)
is proportional to $P_{i,4}$, and we can write
\begin{align}
\sum_{i,j=1}^4 &BR(H \to \eta_i\, \eta_j \to 4\,\tau) =
BR(H \to A A) \times R\; , \nn\\
&R=\left[\sum_{i=1}^4 P_{i,4}^2 \times BR(\eta_i \to
\tau^+\,\tau^-)\right]^2\; .
\label{eq:7}
\end{align}
We can compute $R$ as function of $M_A$ and $X_d$, and the result is
shown in Fig.~\ref{fig:2}. 

In Fig.~\ref{fig:2} we also show  upper bounds on $X_d$ from CLEO (red),
BABAR (blue) and from the condition that one eigenstate of the mass
matrix (\ref{eq:2}) has a mass of 9391~MeV \emph{and} $m_{\eta_{b}(1S)}$
(before mixing) is within a range $9360 - 9445$~MeV covered by QCD
predictions (green). Hence, for $M_A \lsim 10.1$~GeV, only small
values of $X_d$, where $R$ is nearly independent from $X_d$ (as
explained above), are of interest. Like the $BR(\eta_i \to
\tau^+\,\tau^-)$, $R$ varies strongly with $M_A$. It follows from $R
\sim \sum_{i=1}^4 P_{i,4}^4 \times BR(\eta_i \to \tau^+\,\tau^-)^2$
that $R$ never exceeds 0.4 for $M_A$ in the range $9 - 10.5$~GeV, and $R
\sim \left(\Gamma_A^{\tau\tau}/\Gamma_{\eta_b(nS)}\right)^2$ (which is
tiny) as soon as $M_A$ is near any of the masses $m_{\eta_b(nS)}$. Now
the quantity $\xi^2$ constrained by ALEPH (see Fig.~6 in
\cite{Schael:2010aw}) must be interpreted as $\xi^2 = \xi'^2\times R$,
$\xi'^2 = \frac{\sigma(e^+ e^- \to ZH)}{\sigma_\mathrm{SM}(e^+ e^- \to
ZH)}\times BR(H \to 2\, A)$. It follows that $\xi'^2$ is left
unconstrained at least for $M_H \gsim 98$~GeV and $M_A$ in the range $9
- 10.5$~GeV, as well as for any lower value of $M_H$ as long as $M_A$ is
in the range where $R$ in Fig.~\ref{fig:2} is below 0.2, corresponding
essentially to a $BR(\eta_i \to \tau^+\,\tau^-)$ in Fig.~\ref{fig:1}
below $\sim 0.5$ (but depending slightly on $X_d$). Since, in addition,
one always has $\xi'^2 \lsim 1$ even if the process $H \to 2\, A$ is
kinematically allowed (since the $BR(H \to b\,\bar{b})$ is never exactly
zero), scenarios with $M_H \lsim 98$~GeV are consistent
with the ALEPH constraints as well for most values of $M_A$ in the range
$9 - 10.5$~GeV.

\section{Conclusions and outlook}

After the publication of the ALEPH analysis \cite{Schael:2010aw} it
seemed that the attractive scenario with a light CP-even Higgs boson $H$
and a mass $M_H$ well below 114~GeV, decaying dominantly as $H \to 2\ A
\to 4\ \tau$, was tightly constrained. We have shown that these
constraints are absent for $M_H \gsim 98$~GeV and $M_A$ in the range $9
- 10.5$~GeV, and in the case of lower values of $M_H$ for most values of
$M_A$ in this range. The origin is a reduced $BR(A \to \tau^+\,\tau^-)$
caused by $A - \eta_b(nS)$ mixing, leading to dominant hadronic decays
of the physical eigenstates. This window for $M_A$ is of particular
interest, since it contains the region in which the tension between the
observed $\eta_b(1S)$ mass and its prediction based on QCD can be
resolved \cite{Domingo:2009tb, Domingo:2010am} through this mixing. We
emphasize that we did not make particular assumptions on the SU(2)
doublet component $\cos\theta_A$, on $\tan\beta$ or on the coupling
$X_d$ (see (\ref{eq:1})) of $A$ to $b$~quarks since, at least for small
mixing angles, $X_d^2$ cancels out in the expression (\ref{eq:6}) for
the $BR(\eta_i \to \tau^+\,\tau^-)$ for the mass eigenstates.

For small $X_d$ and a correspondingly small $A - \eta_b(nS)$ mixing,
this result seems counterintuitive at first sight. However, the point is
that already a small admixture of any $\eta_b(nS)$ state to the mass
eigenstate $\eta_i$ suffices such that the mass eigenstate $\eta_i$
decays dominantly hadronically, since the corresponding hadronic widths
of $\eta_b(nS)$ are much larger than $\Gamma_A^{\tau\tau}$. This remains
true for small $X_d$, since then $\Gamma_A^{\tau\tau}$ becomes small as
well.

The consequences of this scenario for Higgs searches at the LHC would be
quite dramatic, since the dominant Higgs decay mode would be $H \to 2\ A
\to$~hadrons and, like in the scenarios discussed in
\cite{Carpenter:2007zz, Bellazzini:2009xt,Bellazzini:2009kw}, the $H$
signal would be buried under the QCD background. Moreover, dominant
hadronic decays of the mass eigenstate $\eta_i$ would also handicap
searches for $A$ via central exclusive production \cite{Forshaw:2007ra}
at hadron colliders, or via the $\mu^+\,\mu^-$ final state as proposed
in \cite{Dermisek:2009fd} and studied, using early LHC data, in
\cite{atlas}. It remains to look for $A$ in radiative $\Upsilon$ decays,
but corresponding searches have also to be interpreted carefully taking
mixing effects into account \cite{Drees:1989du,Fullana:2007uq,
Domingo:2008rr, Domingo:2010am}.

\section*{Acknowledgements}

The work of F.D. was supported by the BMBF grant 05H09VKF.


\end{document}